\documentclass[aps,prl,preprint]{revtex4}
\usepackage{color}
\usepackage{epsfig} 
\usepackage{bm}
\pdfoutput=1
\begin{document}

\title{Rashba Spin-Orbit Anisotropy and the Electric Field Control of Magnetism}

\author{ Stewart E.  Barnes$^{1,2}$}
\altaffiliation {barnes@physics.miami.edu}

\author{ Jun'ichi Ieda$^{2,3}$}

\author{Sadamichi Maekawa$^{2,3}$}

\affiliation{
 $^{1}$Physics
Department, University of Miami, Coral Gables, FL 33124, USA. \\
$^{2}$Advanced Science Research Center, Japan Atomic Energy Agency, Tokai, Ibaraki 319-1195, Japan. \\
$^{3}$CREST,Japan Science and Technology Agency, Sanbancho, Tokyo 102-0075, Japan.
}

\hfill

\date{\today}

\begin{abstract}
The control of the magnetism of ultra-thin ferromagnetic layers using an electric field rather than a current, if large enough, would lead to many technologically important applications. To date, while it is usually assumed the changes in the magnetic anisotropy, leading to such a control, arises from surface charge doping of the magnetic layer,
a number of key experiments cannot be understood within such a scenario.
Much studied is the fact that, for non-magnetic metals or semi-conductors,  a large surface electric field gives rise to a Rashba  spin-orbit coupling which leads to a spin-splitting of  the conduction electrons. For a magnet, this splitting is modified by the 
exchange field resulting in a large magnetic anisotropy energy via the Dzyaloshinskii-Moriya mechanism. This different, yet traditional, path to an electrically induced anisotropy energy can explain the electric field, thickness, and material dependence reported in  many experiments. 
\end{abstract}

\maketitle

The possibility of controlling the magnetic anisotropy of thin ferromagnetic films using a static electric field $\bm{E}$ is of great interest since it can potentially lead to magnetic random access memory (MRAM) devices which require less energy than spin-torque-transfer random access memory STT-MRAM\cite{Eerenstein,Chiba2003,Chiba, Stolichnov, Chu,Weisheit,Evgeny}.
Thin magnetic films with a perpendicular magnetic anisotropy (PMA) are important for applications\cite{Emori,Ryu}. 
That an interfacial {\it internal\/} electric field might be used to engineer such a PMA is also of great interest. Experiment\cite{Maruyama,Wang} has indeed shown that such a PMA might, in turn, be modified  by an {\it externally applied\/} electric field,
however the data is usually interpreted in terms of changes to the electronic contribution to magnetic anisotropy due to the surface doping induced by the applied electric field\cite{Chiba,Maruyama,Fowley}.

The theory of the field-induced changes of the magnetic anisotropy reflecting surface doping is invariably developed in terms of  band theory\cite{Kyuno,Tsujikawa,Duan,Niranjan,Nakamura}. The results for both the bulk and thin films can be adequately understood in terms of second order perturbation theory\cite{Freeman1} in which the matrix elements of the spin-orbit interaction are between full and empty states. Large contributions  come from regions where different $d$-bands (almost) cross. That such crossings should be close to the Fermi surface leads to the strong doping dependence in such theories. Nakamura {\it et al.}\cite{Kohji} pointed out a strong negative applied field dependence of the PMA for an isolated mono-layer of Fe(001) arises directly from band splitting rather than doping. Here, for Fe, an $\bm{E}$ perpendicular to the film  breaks reflection symmetry causing a large spin-orbit splitting of $d$-levels near the Fermi surface. 
Despite these important theoretical developments, a clear explanation of a number of key experiments is still lacking.  

Here we develop a simple analytic theory for the existence and electrical control of the PMA based upon the Rashba spin-orbit interaction\cite{Rashba,Casella,Bychkov}
and the single band Stoner model of magnetism. 
We exhibit the somewhat delicate, but very interesting, competition between the Rashba spin-orbit fields and the exchange interaction, reflecting electron correlations. 
This theory can potentially lead to a very large magnetic anisotropy arising from the internal electric fields $E_\mathrm{int}$ which exist at, e.g., ferromagnetic/metal and ferromagnetic/oxide insulator interfaces but modified by the addition of an applied electric field $E_\mathrm{ext}$. There is a  Rashba splitting of the band structure leading to a quadratic, ${E_\mathrm{ext}}^2$, contribution to the magnetic anisotropy, contrasting with a linear in $E_\mathrm{ext}$ doping effect.

 \begin{figure}[b]
\begin{center}
\includegraphics[width=8cm]{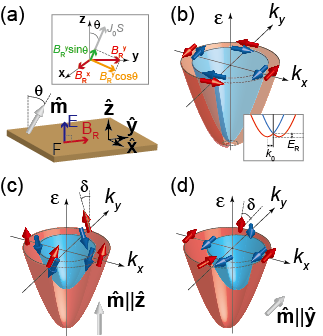}
\end{center}
\caption{(a) The electric field $\bm{E} = E {\hat {\bf z}} $ is perpendicular to the ferromagnet surface while the order parameter direction $ {\hat {\bf m}}$, is defined by the angle $\theta$ relative to ${\hat {\bf z}} $.
 Whatever the  direction of $\bm{k}$, the  Rashba  magnetic field $\bm{B}_R$ of direction $ \bm{k} \times\bm{E}$ lies in the ${\hat {\bf x}}$--${\hat {\bf y}}$ plane. (b) The Rashba split bands of a non-magnetic metal. The two Fermi sheets emerge from a ``Dirac point" near the bottom of the illustration. For the magnetic case the two Fermi sheets are disconnected. (c) For a perpendicular ${\hat {\bf m}} $ the electron spins make a constant  angle $\delta$ to the vertical such that the projection is as in (b). The additional exchange splitting increases as $E^2$. (d) Same but for ${\hat {\bf m}} $ parallel to the plane. With ${\hat {\bf m}} $ along the $y$-direction the majority and minority Fermi seas shift along the $x$-axis in opposite directions. The tilt of the spin relative to ${\hat {\bf m}} $ is no longer a constant being zero along the $x$-axis and a maximum along the $y$-axis. 
}
\label {F100}
\end{figure}

\section{Results}
\noindent\textbf{Model.}
This comprises  a band Stoner model with the Rashba interaction added\cite{Manchon}:
\begin{equation}
\label{100}
H = \frac{p^2}{2m} 
- J_0\bm{S} \cdot \bm{\sigma} + \frac{\alpha_R}{\hbar}  (\sigma_x p_y - \sigma_y p_x),
\end{equation}
where $\bm{p}$ is the electron momentum operator, 
$\bm{S}$ the order parameter, $\bm{ \sigma}$ the Pauli matrices and  $ \alpha_R = e \eta_{\rm so} E$ is the Rashba parameter proportional to $\eta_{\rm so} $, which characterises the spin-orbit coupling, and the magnitude $E$ of the electric field  $\bm{E} = E {\hat {\bf z}}$, taken to be  perpendicular to the plane of the system, and ${\hat {\bf m}}  = \bm{S}/S$ is perpendicular to ${\hat {\bf x}}$ and makes an angle $\theta$ to the ${\hat {\bf z}}$-direction, as in Fig.~\ref{F100}(a). 
\newline

\noindent\textbf{Illustrative non-magnetic example.}  Consider  the Rashba effect in non-magnetic  two dimensional electron gases or surface states on noble metals, e.g., a surface state of Au. As shown in the {\bf methods}, the single particle energy 
\begin{equation}
\label{1001}
\epsilon_{\bm{k}\sigma } = \frac{\hbar^2}{2m} \left(k - \sigma k_0 \right)^2 -  E_R,
\end{equation} 
where $\sigma = \pm 1$ for majority/minority electrons,  the momentum shift $k_0  =  m \alpha_R /\hbar^2$, and 
\begin{equation}
\label{10011}
E_R =  \frac{m {\alpha_R}^2 }{2\hbar^2} 
=  \frac{1}{2}  \left( \frac{e \eta_{\rm so}}{\hbar} \right)^2mE^2, 
\end{equation}
identified in the inset of Fig.~\ref{F100}(b), reflects the single particle energy gain relative to zero electric field, i.e., $E=0$ and $\alpha_R=0$.  For the surface state of Au,  $E_R \approx 3.5$meV,  exemplifying  the energy scale. For the three dimensional problem there is no equivalent shift in $k_z$.

There are many  experiments \cite{LaShell,Hoesch,Ast,Gierz,Krupin} which put in evidence the Rashba splitting in two dimensional electron gases, surface states of noble metals,  bulk layered systems, and e.g., of a surface state of ferromagnetic Yb. 
\newline

\noindent\textbf{Magnetic case - 
origin of the magnetic anisotropy energy.} In the {\bf methods} it is shown the $\theta$ dependent  single particle energy,
\begin{eqnarray}
\label{1006}
\epsilon_{\bm{k}\sigma} & = &
\frac{\hbar^2}{2m}\left[\left(k_{x} -\sigma k_0 \sin \theta\right)^2 + {k_{y}}^2 
\right] -   E_R\sin^2 \theta\nonumber  \\
&& - \sigma \left[
(J_0 S)^2 + \alpha_R^2(  {k_{x} }^2\cos^2\theta + {k_{y}}^2)
\right]^{1/2}.
\end{eqnarray}
The direction of the momentum shift depends upon $\sigma = \pm1$, i.e., the majority/minority character of the band. These shifts also change sign with $ {\hat {\bf m}}  \to -  {\hat {\bf m}} $ for a given $\sigma$. This ``magnetic Rashba splitting" with  $ {\hat {\bf m}}  \to -  {\hat {\bf m}} $ is observed for the surface state of Yb\, \cite{Krupin}.

The contributions to the magnetic anisotropy are highlighted by contrasting  the  perpendicular and parallel orientations of order parameter ${\hat {\bf m}}$ to the plane. 
With ${\hat {\bf m}}$ perpendicular to the plane, i.e., 
${\hat {\bf m}} = {\hat {\bf z}}$ ($\theta =0$), 
the exchange and Rashba fields (see {\bf Methods}) are orthogonal 
and hence the net energy for a single electron Eq.~(\ref{1006}) is 
\begin{eqnarray}
\label{1003}
\epsilon_{\bm{k}\sigma} &=& \frac{\hbar^2}{2m}k^2  
- \sigma   [(J_0S)^2 + (\alpha_R k^2)]^{1/2}.
\end{eqnarray}
The axis of quantisation is tilted by $\delta(k) = \tan^{-1} \frac{\alpha_Rk}{J_0S}$ away from the $z$-axis as shown in Fig.~\ref{F100}(c). 
The majority (minority)  electrons  gain (lose) an energy that is even in $E$. 
This arises from the competition of the Rashba field, perpendicular to ${\hat {\bf m}}$, with the exchange field. 
Such a competition generates  a second order in $E$ contribution to the magnetic anisotropy and is identified with the Dzyaloshinskii-Moriya (DM) mechanism\cite{Dzyaloshinskii, Moriya, book1}.

Now take ${\hat {\bf m}}$  parallel to the $y$-axis, i.e., ${\hat {\bf m}} = {\hat {\bf y}}$ ($\theta =\pi/2$). The $y$-component of $\bm{B}_R$ is parallel to the exchange field and is combined with the kinetic energy. 
The Fermi sea is shifted along the $x$-axis and lowered by $E_R$ as shown in Fig.~\ref{F100}(d).  
This energy gain corresponds to a pseudo-dipolar (PD) contribution to anisotropy energy\cite{book1} which favours an in-plane magnetisation. 
On the other hand, the $x$-component of $\bm{B}_R$, which is perpendicular to $J_0 S{\hat {\bf m}}$, gives rise to a correction to the effective exchange field. 
The direction of the moment tilts away from the $y$-axis in the direction perpendicular to the wave vector by
$\delta(k_y) = \tan^{-1} \frac{ \alpha_R k_y}{J_0S}$ as shown in Fig.~\ref{F100}(d). The  single particle energy, Eq.~(\ref{1006}), is now,
\begin{eqnarray}
\label{1004}
\epsilon_{\bm{k}\sigma} 
&=& \frac{\hbar^2}{2m}[(k_x - \sigma k_0 )^2 +  {k_y}^2]  - E_R\nonumber  \\
&&-  \sigma   [(J_0S)^2 + (\alpha_R {k_y})^2 ]^{1/2},
\end{eqnarray}
where the shift $k_0 $ is the same as in Eq.~(\ref{1001}) but  only along the $x$-axis.   

The effective exchange field 
in Eq.~(\ref{1004}) is smaller than 
that in Eq.~(\ref{1003}) due to the absence of a ${k_x}^2$ term.
This indicates that the overall DM contribution favours a perpendicular ${\hat {\bf m}}$ while the PD term favours an in-plane ${\hat {\bf m}}$. 
This exchange field changes sign for the majority/minority spins, i.e., with $\sigma$.
Assuming  $(J_0 S )^2 >  (\alpha_Rk_x)^2$ and retaining the $\theta$-dependent terms up to the order of $E^2$ in (\ref{1006}), we obtained 
our principal result: 
\begin{equation}
\label{400}
E_{\rm an} =E_R \left[ 1 -  \frac{2T}{J_0 S} 
\right] \cos^2 \theta,
\end{equation}
for the magnetic anisotropy energy, with
\begin{equation}  
\label{T}
T =  \frac{\hbar^2 }{2m} \left(\langle {k_x}^2\rangle_{\uparrow} - \langle {k_x}^2\rangle_{\downarrow}\right),
\end{equation}
where $\langle$ $\rangle$ denotes an  average over the Fermi sea (see {\bf methods}). 
The Rashba  spin-orbit interaction produces a uni-axial anisotropy energy which, as in the Dzyaloshinskii-Moriya theory\cite{book1},  comprises  a direct  second order in $E$ easy plane pseudo-dipolar interaction and an indirect contribution proportional to $ E^2/J_0 S$ 
reflecting the competition between the first order in $E$, Rashba-Dzyaloshinskii-Moriya, and  exchange fields. Clearly an $E^2$ dependent PMA results if $T>J_0S/2$, 
which is the case for real 3$d$ ferromagnets as argued below.
\newline

\noindent\textbf{Competition between the Dzyaloshinskii-Moriya and pseudo-dipolar contributions.}
Taken literally, the Stoner model Eq.~(\ref{100}), with its 
quadratic dispersion, predicts the ratio of  the DM and PD contributions to the PMA. 
The result,  (see {\bf methods}),  depends upon the spatial dimension. In two dimensions the PD and DM terms  cancel although higher order terms ($O({\alpha_R}^4)$) lead to a PMA while in three dimensions the DM term is $-(4/5) E_R\cos^2\theta$ and an in-plane magnetisation is favoured. Lastly, a two dimensional system with a highly anisotropic conductivity might be modelled as a series of parallel one dimensional chains. For chains DM contribution is $-(4/3) E_R\cos^2\theta$ which dominates the PD energy $E_R\cos^2\theta$, appropriate when ${\hat {\bf m}}$ is in-plane and perpendicular to the chains.  Corresponding to the hardest axis, when ${\hat {\bf m}}$ in-plane but rather parallel to the chains, there is neither a DM or PD contribution to the magnetic anisotropy energy. There is thereby a predicted  electric field dependence of the  in-plane anisotropy as seen in early experiments\cite{Chiba}, given the large  compressive strain that arises in  these experiments.

\begin{figure*}[b]
\begin{center}
\includegraphics[width=15cm]{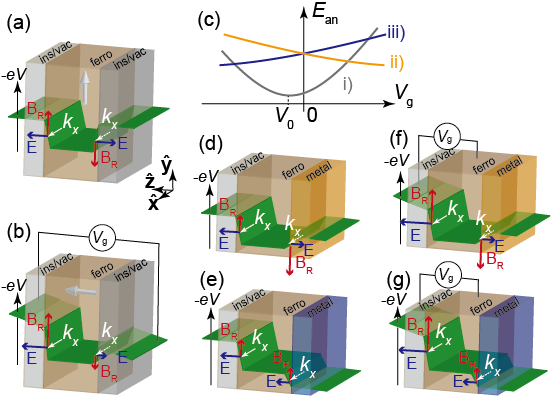}
\end{center}
\caption{(a) There is an electric field $\bm{E}$ in the surface region of a ferromagnet, however for a given wave vector $\bm{k}$, the Rashba field $\bm{B}_R$, proportional to $\bm{k} \times \bm{E}$, has an opposite sign at the two surfaces and the average field is zero. (b) With a finite external field this symmetry is broken and there is a net Rashba  field acting upon the electrons. (c) The gate voltage dependence of the anisotropy energy. The internal electric field causes the shift of the parabola in the lateral axis as indicated by $V_0$ for case i). For cases ii) and iii) the internal field shift is far beyond the external field range and nearly linear $E$-dependence arises. (d) The symmetry is also broken for a insulator-ferromagnet-metal sandwich. Also despite the electric field being smaller at the right surface, for a suitable metal, the spin-orbit coupling is larger and hence the metal interface can still dominate the net Rashba field. (e) Here the work function is larger for the metal than for the ferromagnet and the field for that surface is reversed. Now the Rashba  fields at the two surfaces add. (f), (g) Adding an applied field increases the Rashba  field at the insulating surface which, for this case, causes a net decrease/increase in the average Rashba  field. 
}
\label {F200}
\end{figure*}

However, for the real problem of 3$d$ magnets,  a quadratic dispersion is not at all realistic and the crystal potential $V(\bm{r})$ must be accounted for, see {\bf methods}. For 3$d$ elements $\psi $ is  well localised within the atomic sphere and the averages $\hbar^2\langle {k_x}^2\rangle $, and hence $T$,  are very much  increased as compared to the above na\"{\i}ve estimates. In reality, the DM contribution will invariably lead to a PMA.

\section{Discussion}

The resulting anisotropy energy can be very large. Experiment\cite{LaShell,Hoesch} on conducting but non-magnetic materials helps set the scale. In particular the value of the scaling pre-factor $E_R$ in Eq.~(\ref{400}) for the surface state of Au is $\sim 3.5$\,meV or about 35T in magnetic field units and very much larger than the typical $\sim 1$T  demagnetising field. If a Au film is polarised by contact with an ultra-thin ferromagnet the second factor, $2T/J_0 S$, in Eq.~(\ref{400}) for the field inside a Au surface layer  can be quite large $\sim 5$ leading to a PMA and indeed ultra-thin Fe on Au does have such a PMA\cite{Chagpert,Leeb}. Ultra-thin ferromagnetic films in contact with, e.g., Pt, Pd, and Ta etc., also are found to have a PMA\cite{Fowley,Weisheit}.

Schematically shown in Fig.~\ref{F200}(a) is the potential seen by electrons in a free standing ultra-thin ferromagnetic film. At the surface, the potential reaches the vacuum level within a few atomic spacings. This results in a finite large electric field $E  \sim 10$V/nm at each surface but in opposite senses. Assuming an appreciable  spin-orbit coupling in the interface region, this results, in turn, in a  Rashba  field $\bm{B}_R$ which also changes sign between the two surfaces for a given momentum. Thus for  a perfectly symmetric film, the ferromagnetically polarised electrons see no average field $\bm{B}_R$. This symmetry can be broken by the application of an external electric field as shown in Fig.~\ref{F200}(b).  The electric field is increased at one surface and decreased at the other doubling the net effect. In contrast, for this same symmetric  situation, the surface charges are opposite and doping effects must cancel. Experimentally applied fields of 1V/nm are relatively easy to achieve implying a $\sim10$\% change in the surface anisotropy. Experiments\cite{Nozaki} with a 1.5nm Fe$_{80}$B$_{20}$ sandwiched between two MgO layers is perhaps closest to this situation although  the thickness 1.5nm and 2.5nm of these layers are not equal. Roughly consistent with our estimate [see Fig.~\ref{F200}(c), case(i)], there is \cite{Nozaki} an $\sim 15\%$ symmetric contribution to the magnetic anisotropy for an applied voltage of 2V.

Clearly the intrinsic Rashba  field $\bm{B}_R$  is modified when the material adjacent to the ferromagnets (F), say Fe, are different. In many experiments an insulator $I$, often MgO, lies to one side and a normal metal (N), and e.g., Au, Pt or Pd, completes a tri-layer system. The potential, Figs.~\ref{F200}(d) and (e),  will increase in passing from Fe to MgO but, depending on the effective work-functions, can either increase (Au), Fig.~\ref{F200}(d),  or decrease (Pt), Fig.~\ref{F200}(e), at the FN interface. The latter case is particularly favourable since the intrinsic Rashba  fields have the same sense and add. It is the case that both Pt and Pd N-layers can induce a PMA\cite{Fowley}. Assuming the system is gated on the insulating side and that the PMA is principally due to the N-layer, Figs.~\ref{F200}(f), (g),  the sign of the effect distinguishes between the two cases. If $E$ is increased at the FI interface, the average Rashba field decreases in the first case [Fig.~\ref{F200}(f)] when the effects of the surfaces tend to cancel and, as illustrated in Fig~\ref{F200}(g), increases in the second case when the inverse is true [see Fig.~\ref{F200}(c), case (ii) and (iii)]. Experiment\cite{Fowley}
 indeed shows an opposite field dependence for such systems with Pd and Pt N-layers. That the sign of the electric field contribution to the PMA reflects  the N-layer whereas the field is applied to the opposite surface between F and I supports the current Rashba  model. This is  in stark contrast with the popular surface doping model\cite{Chiba,Fowley}, for which the effects of surface doping are limited by the (possibly magnetically modified) Fermi-Thomas screening length. In reality\cite{Fowley}, the screening length is estimated to be much less than 1nm,  and much too short for there to be an appreciable doping effect of the Pt or Pd layers that are typically distant by a few nanometers. 

Simulating a large applied electric field $E$, the required asymmetry might be controlled in NFN tri-layers by varying in a systematic manner, at the mono-layer  level, the thickness of one of the normal metal layers and by using metals with different spin-orbit couplings. In reality the effect of the substrate transmitted to, and through, the bottom normal metal will imply an asymmetry even for large N-layer thickness. Indeed the PMA surface term for Au/Fe(110)/Au(111) structures does show an non-monotonic dependence on the top Au layer thickness\cite{Chagpert}. 
Experiments\cite{Leeb} for Fe layers on vicinal Ag(001) and Au(001) surfaces and which undergo a symmetry breaking (5$\times$20) surface reconstruction manifest  an in-plane surface term reflecting this broken symmetry and which is larger for Au, with its stronger spin-orbit coupling,  than for Ag. 

It is predicted that the surface coercivity field $H_c$ is proportional to $(E_\mathrm{int} + E_\mathrm{ext})^2$ where $E_\mathrm{int}$ is the internal electric field corresponding to the 
zero-bias Rashba contribution to the anisotropy. Such a non-linear field dependence is observed, e.g., for the in-plane contribution for a (Ge,Mn)As/ZrO$_2$ surface\cite{Chiba}. In other experiments\cite{Wang}  with  CoFeB/MgO/CoFeB structures there is qualitative difference between the $E$ dependence of the anisotropy field $H_c$ of the, ``top'' and ``bottom'', CoFeB layers of this three layer structure, even when they have similar thicknesses. The bottom layer has a larger $H_c$ and is roughly linear while $H_c$ becomes highly non-linear as $H_c \to 0$ as would be expected as $E_\mathrm{ext} \to - E_\mathrm{int}$. 

The most direct  experimental test of the model is the  observation of the band splittings for a model Rashba  system with a variable contact with an itinerant  ferromagnet. This can result in giant magnetic anisotropy (GMA) energies. For example a $E_R\sim 100$meV (or $\sim 1000$T) is reported in angle-resolved photoemission spectroscopy (ARPES) measurements\cite{Ishizaka}  on bulk BiTeI. 
For a thin film of this, or similar material, in contact with an itinerant ferromagnet such as Fe, a suitable exchange splitting $J_0S$, tuned to the order of $E_R$, might be induced and a GMA will result. ARPES performed as a function of the direction of the magnetisation $\bm{m}$ might determine both $E_R$ and the momentum dependence of the exchange splitting leading to estimates of both the PM and DM contributions and which might be directly compared with magnetisation and magnetic resonance measurements. The electrical control of such a GMA has evident important application for non-volatile memory applications. There are clearly many more complicated embodiments of such a device.

In conclusion, it is suggested that the Rashba magnetic field due to the internal electric field in the surface region of an ultra-thin ferromagnet can make an important contribution to the perpendicular magnetic anisotropy. Such surface fields might be modified by application of an applied electric field. Since the internal fields at two surfaces tend to cancel, an asymmetry between the surfaces is important. Such an asymmetry is caused by different metal and insulator caping layers. These ideas are consistent with a large number of experiments. 

\section{Methods}
\noindent\textbf{The non-magnetic case.} This corresponds to Eq.~(\ref{100}) with $J_0=0$. It is solved by taking the axis of quantisation ${\hat {\bf z}}\times \bm{k}$ to be perpendicular to the in-plane $\bm{k}$ as in Fig.~\ref{F100}(b).  
The eigenstates are $e^{i\bm{k} \cdot \bm{r}}|s\rangle$ and $H = (\hbar^2/2m)k^2 - g\mu_B\bm{B}_R \cdot \bm{\sigma}/2$, where the Rashba magnetic field in energy units is defined as $g \mu_B \bm{B}_R = 2\alpha_R(-k_y {\hat {\bf x}}+ k_x {\hat {\bf y}})$, with $\mu_B$ the Bohr magneton and $g$ the g-factor, leaving the spin state $|s\rangle$ to be determined. There  are two  concentric Fermi surfaces.  The energy splitting  $ 2 \alpha_R k  \equiv  \Delta (k/k_F) $, where $\Delta$ is the value for  $k_F \equiv (k_{F\uparrow}+k_{F\downarrow})/2$, with $k_{F\uparrow,\downarrow}$ the Fermi wave number for the majority/minority band. For the surface state of Au, $\Delta \approx 110$meV while $E_F  \approx 420$meV giving the  $E_R\approx 3.5$meV cited in the text. 

\noindent\textbf{The magnetic case.} The full Eq.~(\ref{100})  is solved by defining axes  such that ${\hat {\bf m}}\equiv \bm{S}/S$  lies in the $y$--$z$-plane and $\bm{S} = S(\cos \theta {\hat {\bf z}} + \sin \theta  {\hat {\bf y}})$.
The total field, which defines the axis of quantisation,  $g \mu_B \bm{B}_T =2[(J_0 S + \alpha_R k_x \sin \theta ) {\hat {\bf m}}  -  \alpha_R k_y {\hat {\bf x}} +  \alpha_R k_x \cos \theta  ( {\hat {\bf m}}\times {\hat {\bf x}}  )]$.
It is assumed that,  for a ferromagnet $g \mu_B B_R  < J_0 S$, i.e., the Rashba is smaller than the exchange splitting. To second order in $g \mu_B B_R $, 
$g \mu_B \bm{B}_T \approx 2(J S + \alpha_R k_x \sin \theta ) {\hat {\bf m}}^\prime $ where $JS =  \left[(J_0 S)^2 + \alpha_R^2(  {k_{x} }^2\cos^2\theta + {k_{y}}^2)\right]^{1/2}$ and where ${\hat {\bf m}}^\prime$ differs in direction from ${\hat {\bf m}}$ by a small angle $\delta$ where $\tan \delta \approx \alpha_R(  {k_{x} }^2\cos^2\theta + {k_{y}}^2)^{1/2}/J_0S$. The linear in $k_x$ term, $\alpha_R k_x \sin \theta$,  causes a shift in Fermi sea to give the 
the single particle energy Eq.~(\ref{1006}).


\noindent\textbf{Evaluation of the Dzyaloshinskii-Moriya and pseudo-dipolar contributions.} Needed for $T$ in Eq.~(\ref{400}) are the Fermi sea averages $\langle {k_x}^2\rangle_{\uparrow,\downarrow}=k_{F\uparrow,\downarrow}^2/3$, $k_{F\uparrow,\downarrow}^2/4$, and $k_{F\uparrow,\downarrow}^2/5$,  determined analytically, for quadratic dispersion,  in one, two, and three dimensions respectively. For an isotropic system these averages are 
 related to $J_0 S$ via
\begin{equation}  
\label{4001}
\frac{\hbar^2}{2m} (k_{F\uparrow}^2 - k_{F\downarrow}^2 ) \simeq 2 J_0 S.
\end{equation}
which determines the ratio $2T/J_0S$ in the principal result Eq.~(\ref{400}) given in the text.


\noindent\textbf{Role of the crystal potential.} The effects of the crystal potential $V(\vec r)$ are exhibited by considering a  wave function $\psi = \sum_{\bm{K}} a_{\bm{K}} e^{i(\bm{k} + \bm{K})\cdot \bm{r}}$ which is a linear combination of plane waves, where, $\bm{K}$ are the reciprocal lattice vectors and the $a_{\bm{K}}$ are determined by $V(\bm{r})$. While not convenient for 3{\it d} electrons, at  least in principle, such an expansion in the  true, rather than crystal,  momentum states is always possible. The PD  contribution, $E_R\cos^2\theta$ is independent of the momentum $\bm{k} + \bm{K}$. However  $\hbar^2 \langle {k_x}^2\rangle = \sum_{\bm{K}} |a_{\bm{K}}|^2 \langle (p_x+ \hbar K_x)^2\rangle_\mathrm{BZ}$, where $\langle$ $\rangle_\mathrm{BZ}$ is the average over the first Brillouin zone. For 3{\it d} electrons, the  average  $\hbar^2\langle {k_x}^2\rangle $, and hence $T$, are dominated by the  $a_{\bm{K}} $ for largish $\bm{K}$. It follows  $T$  is enormously increased with the consequences discussed in the text.


\vskip 20pt 

\vfill\eject

%
%

\end{document}